\documentstyle[12pt,titlepage]{article}

\renewcommand{\theequation}{\thesection.\arabic{equation}}

\font\grande=cmr10 scaled \magstep4
\font\medio=cmr10 scaled \magstep2
\outer\def\beginsection#1\par{\medbreak\bigskip
      \message{#1}\leftline{\bf#1}\nobreak\medskip\vskip-\parskip
      \noindent}

\def\laq{\raise 0.4ex\hbox{$<$}\kern -0.8em\lower 0.62 
ex\hbox{$\sim$}}
\def\gaq{\raise 0.4ex\hbox{$>$}\kern -0.7em\lower 0.62 
ex\hbox{$\sim$}}

\begin{document}
\bibliographystyle {unsrt}
\titlepage
\begin{flushright}
DAMTP 96-92\\
UCM 96-04\\
\end{flushright}
\begin{center}{\grande Scalar and Tensor Inhomogeneities}\\
\vspace{5mm} 
{\grande from Dimensional Decoupling\footnote{to appear in
Phys. Rev. D, 55 (1997) }}

\vspace{10mm}
Massimo Giovannini\footnote{
e-mail: M.Giovannini@damtp.cam.ac.uk, giovan@vxcern.cern.ch}\\
{\em Departamento de Fisica Teorica, Universidad Complutense, 
28040 Madrid, Spain} \\ 
and\\
{\em DAMTP, Silver Street, Cambridge, CB3 9EW, U.K.}\\
\end{center}
\centerline{\medio  Abstract} 
We discuss some perturbative techniques  suitable for the 
gauge-invariant treatment of the scalar and tensor inhomogeneities of an 
anisotropic and homogeneous background geometry whose spatial section 
naturally decomposes into the direct product of two maximally symmetric 
Eucledian manifolds, describing a general situation of dimensional decoupling 
in which $d$ external dimensions evolve (in conformal time) with scale factor
$a(\eta)$ and $n$ internal dimensions evolve with scale factor $b(\eta)$.
We analyze the growing mode problem which typically arises in contracting
backgrounds and we focus our attention on the situation where the 
amplitude of the fluctuations not only depends on the external 
space-time but also on the internal spatial coordinates. 
In order to illustrate the possible relevance of this analysis
we compute the gravity waves spectrum produced in some
highly simplified model of cosmological evolution and we find 
that the spectral amplitude, whose magnitude can be constrained by 
the usual bounds applied to the stochastic gravity waves backgrounds, 
depends on  the curvature scale at  which the compactification
occurs and also on the typical frequency of the internal excitations. 
\noindent

\vspace{8mm}

\newpage

\renewcommand{\theequation}{1.\arabic{equation}}
\setcounter{equation}{0}
\section{Introduction}

The assumption of isotropy and homogeneity of the background manifold 
permits a consistent theoretical treatment of the space-time evolution of its
inhomogeneities which can be classified in scalar, vector and tensor modes 
with respect to the three dimensional spatial coordinate transformations
on the constant time hypersurface \cite{1,3}.
The different modes are decoupled to
first order in the amplitude of the fluctuations and this allows the definition 
of perturbed quantities which are invariant under the gauge group of the
infinitesimal coordinates transformations \cite{3,9,9b}.
Within the Bardeen's gauge-invariant approach, the amplified primordial
spectrum of fluctuations can be reliably computed for a wide class of
homogeneous and isotropic cosmological models \cite{4,9} and in particular
for the ``slow-rolling" scenarios leading to the de-Sitter like inflation
\cite{5}.

One of the main motivations in order to relax the assumption of isotropy of the
background geometry comes from the  models  of early universe
  (like superstring theories \cite{6}) describing the unification
of gravity with gauge interactions in a higher dimensional manifold \cite{7}.
We will then consider a homogeneous and anisotropic manifold which can be
written as:
\begin{equation}
ds^2 =g_{\mu\nu}dx^{\mu} dx^{\nu}=
 a^2(\eta) d\eta^2 - a^2(\eta)\gamma_{ij} dx^i dx^j -
b^2(\eta)\gamma_{ab}dy^a dy^{b}
\label{1}
\end{equation}
(conventions: $\mu,~\nu$=$1$,...,$D=d+n+1$; i, j=$1$,...,$d$; 
$a$, $b$=$d$+$1$,...,$n$; $\eta$ is the
conformal time coordinate related, as usual to the cosmic time $t=\int a(\eta)
d\eta$ ; $\gamma_{ij}(x)$, $\gamma_{ab}(y)$ are the metric
tensors of two maximally symmetric Euclidean 
manifolds parameterized,
respectively, by the ``internal" and the ``external" coordinates $\{x^i\}$ and
$\{y^a\}$).
This metric describes the situation in which the external dimensions (evolving
with scale factor $a(\eta)$) and  the internal ones (contracting with scale
factor $b(\eta)$) are dynamically decoupled from each other. In order to
compare the phenomenological consequences
 of the models formulated with extra dimensions it seems
crucial to correctly compute the amplified spectrum of inhomogeneities but,
unfortunately, the treatment of the fluctuations in an anisotropic manifold
becomes quite cumbersome also because of the natural coupling arising
among scalar vector and tensor modes  to first order in the amplitude
of the fluctuations. The original investigations in the subject \cite{2}
 stressed  that the discussion of the fluctuations in the synchronous
 gauge is complicated also because of the spurious gauge modes
 already present \cite{10} in the isotropic case. The problem of the metric
fluctuations in an anisotropic background geometry was then addressed within 
the Bardeen formalism with two different and complementary approaches
\cite{11,12}. It was actually shown that the gauge invariant quantities can be
constructed not only, separately, in the external and in the internal manifold
\cite{11}, but also over the whole manifold of Bianchi-type I \cite{12}
(gravitational waves in Bianchi-I universes were also discussed in \cite{12b}). 
Following the first of the two mentioned approaches it is possible to
distinguish, from the purely  mathematical point of view, the scalar vector and
tensor modes in each of the two manifolds. Even though this classification will
be  technically very useful it does not necessarily coincide with the
physical situation  (as correctly stressed in \cite{11}), since, for
example, the tensor fluctuations polarized along the internal dimensions will
be seen by an observer living in the external space as scalar fluctuations.
The evolution equations for each type of perturbations were then solved well
outside the horizon under the assumption that the Laplacians belonging to the
external and internal manifolds were negligible. The anisotropic  extension
of the scalar Bardeen potentials were shown to grow much faster than the tensor
and vector gauge-invariant  amplitudes in the vicinity of the collapse of 
the internal scale factor. 
The very fast growth of the scalar modes outside the horizon was also 
discussed in the context of the dilaton-driven 
solutions in string cosmology \cite{13}, where it was
found \cite{14} that even though the rate of increase of the scalar 
fluctuations is much faster than in the usual inflationary models 
characterized by a quasi de-Sitter  spectrum \cite{4} a perturbative 
treatment is still plausible, at least in the $(3+1)$-dimensional case 
with static internal dimensions, by carefully ``gauging-down" the scalar 
growing modes. 
In spite of these attempts, in an anisotropic background the solution of the
evolution equations well outside the horizon does not suffice, by itself, for 
the calculation of the spectrum of metric perturbations. 
In order to give a reliable expression for the space-time evolution of the
proper amplitude of the fluctuations it can be assumed \cite{15} relying on
the particular 
features of the background evolution, that the only effective dependence of
the perturbed quantities from the internal dimensions comes in through the time
evolution of the compactification radii.
This approximation scheme can be illustrated using the evolution equation of
the tensor modes polarized along the external dimensions (which we will derive
in Sec. 3):
\begin{eqnarray}
{h_i^j}'' + [(d-1){\cal H} + n{\cal F}] {h_i^j}' -\nabla^2 _{\bar{x}} {h_i^j} -
\frac{a^2}{b^2} \nabla^2_{\bar{y}} {h_i^j} =0
\label{2}
\end{eqnarray}
($'= \partial/ \partial\eta$; ${\cal H}= (\ln{a})'$, ${\cal F}= (\ln{b})'$;
 $\nabla^2_{\bar{x}}$, and $\nabla^2_{\bar{y}}$ are, respectively,
the external and the internal Laplace-Beltrami operators).
In some particular model it can actually happen that
$a\laq b$, at least for scales
 which went out of the horizon before the compactification was achieved.
In this case it is possible to neglect the internal gradients compared 
to the external ones, and all the
dependence from the internal dimensions will be given by ${\cal F}$ which
vanishes in the case of static internal scale factors.
 If, on
the contrary $b\laq a$, the internal Laplacians cannot be neglected especially
prior to the dimensional decoupling when the internal and external scale
factors were of the same order. 
Since the tensor modes only couple to the background curvature 
in order to discuss their evolution we only need to specify the 
time evolution of the scale factors.
The scenario we want to examine consists in general of two phases. 
A multidimensional phase where the time evolution of the scale factors can be 
parameterized as
\begin{equation}
a(\eta) \sim |\eta|^{\alpha},~~~b(\eta)\sim |\eta|^{\beta}~~.
\label{3}
\end{equation} 
This phase can be genarally followed by a compactification phase which 
glues together the multidimensional epoch and the ordinary, isotropic,
FRW evolution. There are different compactification scenarios corresponding
to the parameterization (\ref{3}).
In It is actually possible either 
to assume that the compactification occurs at some stage after the initial 
``big-bang" singularity (i.e. for $\eta>0$ in (\ref{3})) as, for instance, in 
\cite{7b,18b3}) or at some stage before the ``big-bang"  
singularity (i.e. $\eta<0$ in (\ref{3})) 
as seems more likely in the pre-big-bang
scenarios \cite{13,14b}. Even though our considerations, 
at this stage, are purely kinematical, it is anyway useful to point out that
the two mentioned compactification pictures are dynamically very different
since in the pre-big-bang case the end of the higher dimensional phase 
could coincide, in principle, with the beginning of the ordinary isotropic
evolution while in the first picture some other mechanism is required 
in order to smoothly connect the multidimensional phase (trapped among two
singularities) to the FRW universe.
An interesting issue is then if in the context of the string
inspired models of cosmological evolution the usual problems of the ordinary
Kaluza-Klein models (stabilization and isotropization of 
the internal dimensions, backreaction effects due to particle production 
\cite{9bG,9b1G}) can be solved (or at least alleviated)  by the mechanisms
usually proposed \cite{18} 
in order to regularize the time  evolution of the 
curvature invariants and in order to slow down the dilaton growth.
Since at the moment this solution is still unclear not only in more than
four dimensions but even
in  some simplified two-dimensional toy model of
cosmological evolution \cite{18b2} (where the quantum backreaction was 
taken into account and where the problem of the dilaton seem growth seem 
still to  persist)
we will concentrate our attention on the main kinematical features of the
dimensional decoupling. Our purpose will only be to stress the regimes of the
time evolution of the scale factors
where, possibly, the internal Laplacians of eq. (\ref{2}) are leading if
compared to the internal ones.
More specifically for $t>0$ an eventual accelerated expansion 
in the external space 
($\ddot a>0$, $\dot a>0$, i. e. $\alpha \leq -1$ in (\ref{3})) 
required in order to solve the kinematical difficulties of the 
Standard Model, together with a
simultaneous contracting evolution of the internal dimensions ($\dot b<0$,
$\beta<0$ in (\ref{3})), would not forbid in the large $t$ limit
($|\eta|\rightarrow 0$ if $\alpha \leq -1$) the dominance of the internal
Laplacians over the external ones ($\alpha<\beta$ in our parametrization).
Similar conclusions can be reached in the limit $\eta\rightarrow 0^{-}$ if, for
$t<0$, we consider an accelerated contraction in the external space ($\ddot
a<0, \dot a<0, -1<\alpha<0$) as suggested by the Einstein frame picture of the
string cosmological models \cite{17,18} in order to solve the flatness and the
horizon problems.
These qualitative considerations suggest that if a given external Fourier mode
$k$ went out of the horizon ($k\eta\sim 1$) during an early phase where $a\gaq 
b$, the contribution of the internal Laplacians have to be seriously considered.
At the same time a complete solution of the evolution equations of the scalar
and tensor fluctuations depending on the internal and external spatial
coordinates was never studied not even in some oversimplified model of 
background evolution.
A reliable computation of the power spectrum for scalar
and tensor inhomogeneities in higher dimensional theories  is beyond the scope 
of the present investigation, nonetheless we 
would like to analyze the evolution and the
amplification of the metric fluctuations in some specific toy model with extra
dimensions but without assumptions for what concerns the evolution equations
of the fluctuations. We would like also to avoid any kind of ``slow-rolling"
hypothesis in the solution of the background equations 
 which could confuse the analysis of the perturbations. For this
reason we shall mainly discuss two classes of exact solutions of the
multidimensional Einstein equations : the Kaluza-Klein ``vacuum" solutions 
which 
can represent a good description in the vicinity of the collapse of the
internal dimensions \cite{18b2} and the multidimensional anisotropic universe
filled with  scalar field  matter.
We find quite useful to work from the very beginning with the following
scalar-tensor action:
\begin{equation}
S= S_{g} + S_{m}= -\frac{1}{6 l_D^2} \int d^D x \sqrt{-g}R +
\int d^D x \sqrt{-g}
\{\frac{1}{2}g^{\alpha\beta}\partial_{\alpha}\varphi\partial_{\beta}\varphi -
V(\varphi)\}       
\label{4}
\end{equation}
(where $l_D= \sqrt{8\pi G_D/3}$; if $V=0$ and $\varphi=0$ the action can
describe a vacuum Kaluza-Klein phase and if ${\varphi'}^2 >> a^2 V$ we recover
the tree level string theory effective lagrangian [in $D=10$ critical 
dimensions] for the massless modes of the theory,
written in the Einstein frame and in the absence of antisymmetric tensor
field).

The plan of the paper is the following. 
In Sec. 2  we will review the higher
 dimensional background equations of motion and the particular 
classes of exact solutions which will be used in the 
following sections as theoretical laboratory for the analysis 
of the fluctuations.
In Sec. 3 the Bardeen approach for the scalar and tensor perturbations will be
discussed. Particular attention, in the case of the scalar fluctuations, will
be paid to the possible gauge choices 
which completely fix the coordinate frame
 and to the diagonalization of the system of perturbed equations.
In Sec. 4 we will focus our study  on the evolution of tensor perturbations
and we will compute the normalized spectral amplitude  for two simplified
models of dimensional decoupling.
In Sec.5 we will move to the analysis of the scalar inhomogeneities and we will
approach the growing mode problem within the formalism discussed in the 
previous Sections. Sec. 6 contains few concluding remarks.

\renewcommand{\theequation}{2.\arabic{equation}}
\setcounter{equation}{0}
\section{Background models}

The variation of the action (\ref{4}) with respect to $g_{\mu\nu}$ 
and to $\varphi$ provide the equations of motion for the background fields: 
\begin{equation}
R_\mu^\nu - \frac{1}{2}\delta_{\mu}^\nu R = 3 l_D^2 T_{\mu}^{\nu}
\label{ein}
\end{equation}
\begin{equation}
g^{\alpha\beta}\nabla_{\alpha}\nabla_{\beta}\varphi+ \frac{\partial
V}{\partial\varphi}=0
\label{dil} 
\end{equation}
(where $T_\mu^\nu= \partial_\mu\varphi\partial^\nu \varphi -g_{\mu\nu}
(\frac{1}{2}g^{\alpha\beta}\partial_\alpha\varphi\partial_\beta\varphi-
V(\varphi))$).
If we restrict our attention to the case in which the scalar field
 is homogeneous ($\varphi=\varphi(\eta)$) the evolution of the geometry
 is completely determined by the time evolution of the two scale
factors $a(\eta)$ and $b(\eta)$. 

 Using the line element (\ref{1}), eqs.
(\ref{ein})-(\ref{dil}) become:
\begin{eqnarray}
\frac{d(d-1)}{2}{\cal H}^2+\frac{n(n-1)}{2} {\cal F}^2+ n d {\cal H}
{\cal F}~~~~~~~~~~~~~~~~~~~~~ 
\nonumber\\
+ \frac{d(d-1)}{2} {\cal K}_a 
+\frac{n(n-1)}{2}\frac{a^2}{b^2}{\cal K}_b 
= 3 l_D^2 (\frac{\varphi'^2}{2} +  a^2 V),~~~~~~~~~~(00)
\nonumber
\end{eqnarray}
\begin{eqnarray}
(d-1) {\cal H}'+\frac{(d-1)(d-2)}{2}{\cal H}^2+n {\cal 
F}'+\frac{n(n+1)}{2} {\cal F}^2+ n (d-2){\cal H}{\cal F}
\nonumber\\
+ \frac{(d-2)(d-1)}{2} {\cal K}_a + 
\frac{n(n-1)}{2}\frac{a^2}{b^2}{\cal K}_b 
= 3 l_D^2( a^2 V -\frac{\varphi'^2}{2} )~~~~~~~~(ii)
\nonumber
\end{eqnarray}
\begin{eqnarray}
(n-1){\cal F}'+ d {\cal H}'+\frac{d(d-1)}{2} {\cal H}^2+
\frac{n(n-1)}{2}{\cal F}^2+ (d-1)(n-1) {\cal H}
{\cal F}
\nonumber\\
+ \frac{d(d-1)}{2} {\cal K}_a + 
\frac{(n-2)(n-1)}{2}\frac{a^2}{b^2}{\cal K}_b 
= 3 l_D^2( a^2 V -\frac{\varphi'^2}{2})~~~~~~~~~(aa) 
\nonumber
\end{eqnarray}
\begin{equation}
\varphi''+[(d-1){\cal H} +n {\cal F}]\varphi'+
\frac{\partial V}{\partial\varphi} =0~~~~~~~~~~~~~~~~~~~~~~~~~~~(\varphi)
\label{background}
\end{equation}
(where ${\cal K}_a$ and ${\cal K}_b$ 
are respectively the curvatures constants of the external and internal 
maximally symmetric spaces).
In this paper we will generally work in the case of spatially flat internal 
and external manifold with topology $M_{d+1}\otimes T_n$ (where $d+1$ is the
conventional $(d+1)$-dimensional flat universe and $T_n$ is an $n$-dimensional
torus).
Summing and subtracting the previous equations we get two useful relations:
\begin{eqnarray}
{\cal F}'&=&-{\cal F}[n{\cal F}+(d-1){\cal H}] + \frac{6 l_D^2 a^2 V}{(n+d-1)}
\nonumber\\
{\cal H}'&=&-{\cal H}[n{\cal F}+(d-1){\cal H}] + \frac{6 l_D^2 a^2 V}{(n+d-1)}
~~~~.
\label{third}
\end{eqnarray}
If $V=0$ and $\varphi=0$  the solutions of the system
(\ref{background}) define the Kaluza-Klein vacuum \cite{18b2}. 
A class of exact vacuum
solutions can be obtained using the power law ansatz of eq. (\ref{3}) in the
equations of motion (\ref{background}) 
\begin{eqnarray}
\alpha &=& \frac{d\pm\sqrt{d^2 +d (n+d)(n-1)}}{d(n+d-1)
\mp\sqrt{d^2 + d(n+d)(n-1)}}
\nonumber\\
\beta &=& \frac{nd \mp d\sqrt{d^2 +d(n+d)(n-1)}}{nd(n+d-1) \mp n\sqrt{d^2 + 
d(n+d)(n-1)}}
\label{sol1}
\end{eqnarray}
(the exponents of the scale factors in cosmic time ($\tilde\alpha\equiv
\alpha/(\alpha+1)$, $\tilde\beta\equiv\beta/(\alpha+1)$) are related by the 
Kasner sum rules $d\tilde\alpha+n\tilde\beta=1$ and 
$d{\tilde\alpha}^2+ n{\tilde\beta}^2=1$).
The twofold ambiguity in the sign of the exponents shows that there are two
independent solutions for each number of internal and external dimensions.
A particularly simple case which will be used in our
analysis is the solution with $n=1$ and $d=3$. In this case the two solutions
are $\alpha =0$, $\beta= 1$ and $\alpha=1$, $\beta=-1$. Since we want to analyze
mainly the contribution of the internal dimensions to the evolution of the
fluctuations we choose the solution i which the external dimensions are static
($\alpha=0$, $\beta=1$) so that all the contribution to the amplification of the
fluctuations will come, effectively, from the internal space.
The contracting branch 
($\beta = +1$, i. e. $\ddot b=0$ and $\dot b<0$ since $t \sim \eta$) 
can be matched with a radiation phase
\begin{eqnarray}
a(\eta)= 1 ,~~~~~~b(\eta) =(- \frac{\eta}{\eta_c}),~~~~~\eta\leq -\eta_c
\nonumber\\
a(\eta)=(\frac{\eta+2\eta_c}{\eta_c}),~~~~~b(\eta)=1,~~~~~\eta\geq -\eta_c~~.
\label{toy1}
\end{eqnarray}
This toy model is not realistic and somehow
artificial since the radiation is not dynamically
generated but only assumed. In a more refined treatment 
the back-reaction effects should be correctly taken into account 
\cite{9bG,9dG} since the 
scalar and tensor inhomogeneities amplified during the classical evolution
can eventually modify the background dynamics leading to an effective damping of
the anisotropy  of the  background metric (as usually happens in the 
($3+1$)-dimensional anisotropic models of Bianchi type-I \cite{9cG}).
Nonetheless (\ref{toy1}) shares some essential features of a realistic scenario
of dimensional reduction in which the $(3+1)$ external
dimensions decouple from the fifth one down to a compactification scale $H_c\sim
1/\eta_c$ and our purpose will be to connect the amplitude of the scalar and
tensor fluctuations not only with the curvature scale but also with the typical
frequency of the internal oscillations (which can be also constrained, with
different arguments, from the present value of the fine structure
constant\cite{7b}).
 
In the case of negligible potential, using the previous power-law
ansatz for the scale factors from eq. (\ref{2}) and a logarithmic ansatz for the
scalar field
\begin{equation}
\varphi \sim \frac{\gamma}{\sqrt{3l_D^2}}\ln{|\eta|}
\end{equation}
another class of solutions of eq. (\ref{background}) is given by:
\begin{eqnarray}
\alpha &=& -\frac{\mp (1-d-n) \pm n \mp (d-1) -\sqrt{d+n} -1}{(\sqrt{d+n}
+1)(d+n-1)}
\nonumber\\
\beta &=& -\frac{\pm (1-d-n) \pm n \mp (d-1) -\sqrt{d+n} -1}{(\sqrt{d+n}
+1)(d+n-1)} 
\nonumber\\
\gamma &=& 
-\sqrt{\frac{2}{d+n-1}}\frac{\pm n \mp (d-1) -\sqrt{d+n} -1}{(\sqrt{d+n}+1)}~~.
\label{sol2}
\end{eqnarray}
By choosing everywhere the upper sign and in the case 
of critical dimensions ($D=10$ with $d=3$ and $n=6$) 
the solution (\ref{sol2}) is a particular example of the 
general exact dilaton-driven solutions originally
 derived \cite{14}  in the String frame usually related to the Einstein frame 
by a conformal rescaling of the metric tensor
($\tilde{g}_{\mu\nu}^{String}= g_{\mu\nu}^{Einstein} 
\exp{[(2\varphi-2\varphi_1)/(d+n-1)]}$;  
the dilaton is only redefined according to 
$\varphi_{Einstein} = \sqrt{2/(d+n-1)}\tilde{\varphi}_{String}$ 
[$6l_{D}^2 =1$ so that $\varphi$ is dimension-less]).
We will discuss, as a particular  toy model, a $10$-dimensional dilaton-driven
solution continuously matched with the radiation phase: 
\begin{eqnarray}
a(\eta)= (-\frac{\eta}{\eta_c})^{-1/4},~~~~
b(\eta) =(-\frac{\eta}{\eta_c})^{1/4},~~~
\varphi(\eta) = \frac{\sqrt{3}}{8 l_D^2}\ln{(-\frac{\eta}{\eta_c})},~~~~
\eta\leq -\eta_c
\nonumber\\
a(\eta)=(\frac{\eta+2\eta_c}{\eta_c}),~~~~~b(\eta)=1,
~~~~\varphi= const.~,~~~~~\eta\geq -\eta_c~~.
\label{toy2}
\end{eqnarray}
In this model for $\eta<-\eta_c$ we have an accelerated contraction of the
external dimensions  supplemented by the a decelerated contraction of the
internal ones.
We also notice that for $\eta<-\eta_{c}$ the scale factor are related by a
duality relation ($a=1/b$) which more generally  holds in the String frame 
($\tilde{a}= 1/\tilde{b}$) provided we choose the upper sign in (\ref{sol2}).
We point out that this model is not realistic
for the same reasons mentioned in the case of (\ref{toy1}) and also
because it was shown that in order to have a
graceful exit from the dilaton-driven epoch it is crucial to include in the
picture a stringy phase during which the background dynamics is driven by the 
higher order in the string tension expansion \cite{18}.
In both  the examples (\ref{toy1}) and (\ref{toy2}) the internal scale
factors are static during the radiation dominated era, 
while a time
dependence could be, in principle, also included in the internal scale factors 
during the radiation and matter dominated epochs.
 The time dependence in
the internal scale factors would be anyway strongly constrained
 by nucleosynthesis which would require \cite{20,21}
 during the radiation dominated epoch
 $b_{ns}/b_0 < 1+
\epsilon$ ($b_0$ is the actual value of $b$ and $|\epsilon|<10^{-2}$),
while in the matter dominated epoch the constraint
would be instead $\dot b/ b\equiv {\cal F}/ a< 10^{-9} H_0$ 
($H_0= 1.1 \times 10^{-28} h_{100}~ cm^{-1}$). Moreover the time variation 
of $G_D$ is also constrained \cite{22}
 since $G_{D}(nucl)/G_D(\eta_0) =1+\epsilon$
($|\epsilon|< 3\times 10^{-1}$) in the radiation dominated epoch, and
$|\dot{G}_D/ G_D|< 10^{-1} H_{0}$ during the matter dominated epoch. In our
naive models $b=1$ (for $\eta>-\eta_c$) the previous constraints are
automatically satisfied. 

\renewcommand{\theequation}{3.\arabic{equation}}
\setcounter{equation}{0}
\section{Scalar and tensor fluctuations}

The scalar and tensor fluctuations of the geometry (\ref{1}) can be discussed 
within a generalization \cite{11} of the gauge-invariant formalism \cite{3,9}.
The infinitesimal coordinate transformations preserving the scalar nature 
of the fluctuations with respect to each maximally symmetric space are:
\begin{eqnarray}
x^i\rightarrow \tilde{x}^i &=& x^i +  \epsilon^i 
\nonumber\\
y^a\rightarrow \tilde{y}^a &=& y^a + \zeta^a 
\nonumber\\
\eta\rightarrow \tilde{\eta} &=& \eta + \epsilon^0 ~~~~~,
\label{transform}
\end{eqnarray}
(where $\epsilon^0 =\epsilon^0(\overline{x}, \overline{y}, \eta)$,
$\zeta^a\equiv \partial^a \zeta(\overline{x}, \overline{y}, \eta)$,
$\epsilon^i\equiv \partial^i \epsilon(\overline{x}, \overline{y}, \eta)$).
The perturbed scalar metric can be written in terms of $8$ linearly independent
scalar quantities
\begin{equation}
\delta g_{\mu\nu}^{(S)}= \left(\matrix{2 a^2 \phi&- a^2 B_{i}&-ab C_a&\cr
- a^2 B_{i}&2 a^2 \psi\delta_{ij}-2 a^2 E_{ij}&-ab D_{ia}&\cr
-ab C_a&-ab D_{ia}& 2 b^2 \xi\delta_{ab}-2b^2 G_{ab}&\cr}\right)  
\end{equation}
(conventions :$B_{i}=B_{|i}$, $E_{ij}=E_{|ij}$, $C_a =C_{|a}$, $G_{ab}=
G_{|ab}$,
$D_{ia}=D_{|ia}$; the bar denote the covariant derivative with respect to one
of the two
internal spatial metrics depending on  the index, and it coincides with the
ordinary partial derivative if ${\cal K}_a={\cal K}_b=0$).
The fluctuations in the scalar field will be
\begin{equation}
\varphi(\eta,\overline{x}, \overline{y})\rightarrow \varphi(\eta) +\chi(\eta,
\overline{x}, \overline{y})~~~.
\end{equation}
Under an infinitesimal coordinate transformation (\ref{transform}) the
perturbed scalar quantities change as follows:
\begin{eqnarray}
\phi &\rightarrow & \tilde\phi= \phi - {\cal H} \epsilon^0 - {\epsilon^0}'
\nonumber\\
\psi &\rightarrow & \tilde\psi = \psi  + {\cal H} \epsilon^0
\nonumber\\
\xi & \rightarrow & \tilde\xi =\xi + {\cal F}  \epsilon^0
\nonumber\\
E &\rightarrow &\tilde{E}=E- \epsilon 
\nonumber\\
B &\rightarrow &\tilde{B}=B +\epsilon^0 - \epsilon' 
\nonumber\\
C &\rightarrow &\tilde{C}=C + \frac{a}{b}\epsilon^0 - \frac{b}{a}\zeta'
\nonumber\\
D &\rightarrow &\tilde{D}=D -\frac{b}{a}\zeta - \frac{a}{b}\epsilon
\nonumber\\
G &\rightarrow &\tilde{G}=G - \zeta
\nonumber\\
\chi &\rightarrow & \tilde{\chi}= \chi - \varphi' \epsilon^0 ~~~~.
\label{gauge}
\end{eqnarray}
A possible set of linearly independent gauge-invariant quantities is then:
\begin{eqnarray}
\Phi &=& \phi +\frac{1}{a}[(B-E')a]'
\nonumber\\
\Psi &=&\psi +{\cal H}(B-E')
\nonumber\\
\Xi &=&\xi + {\cal F}(B-E')
\nonumber\\
\Omega &=& \frac{a}{b}{\cal F} C - G'{\cal F} + \frac{a^2}{b^2} \xi
\nonumber\\
\Theta &=& D - \frac{b}{a} G - \frac{a}{b} E
\nonumber\\
X &=& \chi +\varphi' (B-E')~~~.
\label{bardeen}
\end{eqnarray}
Notice that $\Psi$ and $\Phi$ coincides, up to a sign, with the Bardeen's
potentials \cite{2} while $\Xi$, $\Omega$ and $\Theta$ appear only in the
anisotropic case.
In the homogeneous and isotropic case it is always possible to choose a
particular coordinate system by completely fixing, to first order, the
arbitrary scalar functions appearing in the transformations (\ref{transform}).
If the scalar functions are completely fixed (like in the case of
the conformally newtonian gauge \cite{3,9}) the equations of motion of
 the fluctuations
will be second order differential equations, if, on the contrary, the
infinitesimal scalar
functions are not completely fixed (like in the case of the synchronous
gauge\cite{1,3,10})
the evolution equations will be of course linear but of higher order. 
Since we want to make our
problem more tractable we completely fix the coordinate system 
\begin{equation}
\tilde{D}=0,~~~\tilde{B}=0,~~~\tilde{E}=0,
\label{pert}
\end{equation}
in eq. (\ref{gauge}) and, as a consequence, $\epsilon^0$, $\epsilon$ and
$\zeta$ are determined from the same equation. In this gauge the longitudinal
fluctuations ($\phi$, $\xi$, $\psi$) coincide with the corresponding
gauge-invariant quantities defined in (\ref{bardeen})
 and in this sense it represents a generalization to
the anisotropic case of the conformally newtonian gauge.
By perturbing to first order the Einstein equation (\ref{ein}) and the 
scalar field equation (\ref{dil}) we obtain:
\begin{equation}
\delta R_{\mu}^{\nu}- \frac{1}{2}\delta_{\mu}^{\nu}
( g^{\alpha\beta}\delta R_{\alpha\beta}
+ \delta g^{\alpha\beta} R_{\alpha\beta})=3 l_{D}^2 \delta T_{\mu}^{\nu}
\label{pertein}
\end{equation}
\begin{eqnarray}
\delta g^{\alpha\beta}(\partial_{\alpha}\partial_{\beta}\varphi-
\Gamma_{\alpha\beta}^{\sigma}\partial_{\sigma}\varphi) +
g^{\alpha\beta}(\partial_{\alpha}\partial_{\beta}\chi+
\delta\Gamma_{\alpha\beta}^{\sigma}\partial_{\sigma}\varphi +
\Gamma_{\alpha\beta}^{\sigma}\partial_{\sigma}\chi)-
\frac{\partial^2 V}{\partial \varphi^2}\chi=0
\end{eqnarray}
($\delta\Gamma_{\alpha\beta}^{\sigma}$ and $\delta R_{\mu\nu}$ are the 
perturbed affine connections and the perturbed Ricci tensors, the indices are
raised using always the background metric
$g_{\mu\nu}$).
By using the background field equations (\ref{ein})-(\ref{dil}) we can write
down explicitly the evolution equations of the fluctuations
 given the perturbed form of the metric
(\ref{pert}) in the gauge (\ref{gauge}).
The ($i\neq j$) component of eq. (\ref{pertein}) implies
\begin{equation}
\phi=(d-2)\psi + n\xi - \nabla^2_{\bar{y}}G
\label{phi}
\end{equation}
which allows to eliminate $\phi$ from all the evolution equations of the
perturbations. From the ($0i$),
($0a$) and ($aj$) components of (\ref{pertein}) we get, respectively: 
\begin{eqnarray}
(d-1)\psi'+(d-2)\psi[(d-1){\cal H} + {\cal F}] + n\xi' 
+n\xi [(n+1) {\cal F} + (d-2) {\cal H}] 
\nonumber\\
-(\nabla^2_{\bar{y}}G)' - [(d-2){\cal H} +(n+1) {\cal F}] \nabla_{\bar{y}}^2 G 
+ \frac{a}{2b} \nabla^2_{\bar{y}}C =3l_D^2 \varphi'\chi~~~~~(0i),
\label{0i}
\end{eqnarray}
\begin{eqnarray}
-\frac{b}{2a}\nabla^2_{\bar{x}} C + (n-1)\xi'+ d \psi'+ [d(d-1){\cal H} +
(d-2)(n-1){\cal F}] \psi 
\nonumber\\
+ ( n\xi - \nabla^2_{\bar{y}}G) (d{\cal H} + (n-1){\cal F})= 3 l_D^2
\varphi'\chi~~~~~~~~~~~~~~~~(0a),
\label{0a}
\end{eqnarray}
\begin{eqnarray}
-\frac{b}{2a} [(d-2) {\cal H} + (n+1) {\cal F}] C - \frac{b}{2a} C'
+ \psi - \xi + \nabla^2_{\bar{y}}G=0~~~~~~~~~(aj),
\label{aj}
\end{eqnarray}
which are not equations of motion but only constraints connecting the
fluctuations to their first time derivative.
The repeated use of eq. (\ref{phi}) together with the equations of motion of
the background fields allows finally to write down the ($00$), ($i=j$), 
($a=b$), ($a\neq b$) components of the evolution equations
\begin{eqnarray}
&-&\frac{1}{2} (\nabla^2_{\bar{y}}G)''
+ \frac{1}{2}(\nabla^2_{\bar{y}}G)'[(n-2){\cal F} + (d+1){\cal H}] 
- \frac{1}{2}\nabla^2_{\bar{x}}\nabla^2_{\bar{y}} G
- \frac{1}{2}\frac{a^2}{b^2}\nabla^2_{\bar{y}}\nabla^2_{\bar{y}} G  
\nonumber\\
&+&n\nabla^2 \xi - \xi'[n d{\cal H}+n(n-1) {\cal F}] +(d-1)\nabla^2 \psi 
- \psi' [d(d-1){\cal H} + n d{\cal F}] 
\nonumber\\
&+&\frac{a}{2b} (\nabla^2_{\bar{y}}C)' 
-\frac{a}{2b}[d{\cal H} 
+ (n-1){\cal F}] \nabla^2_{\bar{y}}C 
\nonumber\\
&=& 3 l_D^2 [{\varphi'\chi'} + 2 a^2 V ((d-2)\psi + n\xi -\nabla_{\bar{y}}^2G
+ a^2 V' \chi)]~~~(00),
\label{00}
\end{eqnarray}
\begin{eqnarray}
&-&\frac{1}{2}(\nabla^2_{\bar{y}}G)''
-\frac{1}{2}(\nabla^2_{\bar{y}} G)'[(3d-5){\cal H} + (3n +2){\cal F}]
- \frac{1}{2}\nabla^2_{\bar{x}}\nabla^2_{\bar{y}}G 
-\frac{1}{2}\frac{a^2}{b^2}\nabla^2_{\bar{x}}\nabla^2_{\bar{y}}G 
\nonumber\\
&+&n\xi''+\xi'[ 2 n (d-1){\cal H} + n(2n+1) {\cal F} -n {\cal H}] 
\nonumber\\
&+&(d-1)\psi''+ \psi' [ (d-1)(2d-3) {\cal H} + n (2d-3) {\cal F}]
\nonumber\\
&+&\frac{a}{2b}(\nabla^2_{\bar{y}}C)'
-\frac{a}{2b}(\nabla^2_{\bar{y}}C) [(d-2){\cal H} + (n+1){\cal F}]
\nonumber\\
&=&3 l_D^2[\varphi'\chi'- a^2 V'\chi - 2a^2 V( (d-2)\psi +n\xi
-\nabla^2_{\bar{y}}G)]~~~(i=j),
\label{ii}
\end{eqnarray}
\begin{eqnarray}
&-&\frac{1}{2}(\nabla^2_{\bar{y}} G)'[(3d-1){\cal H} 
+ (3n -2){\cal F}]
- \frac{1}{2}\nabla^2_{\bar{x}}\nabla^2_{\bar{y}}G 
-\frac{1}{2}\frac{a^2}{b^2}\nabla^2_{\bar{x}}\nabla^2_{\bar{y}}G 
\nonumber\\
&-&\nabla^2 \psi +  d\psi'' + \psi'[2d(d-1) {\cal H} + 2 (d-1)(n-1){\cal F}]
\nonumber\\
&+&\nabla^2 \xi + (n-1)\xi'' +   
\xi'[(d(2n-1)- (n-1)) {\cal H} + 2n(n-1){\cal F}]
\nonumber\\
&+&\frac{a}{2b}(\nabla^2_{\bar{y}}C) [d{\cal H} + (n-1){\cal F}]
+\frac{a}{2b}(\nabla^2_{\bar{y}}C)'
\nonumber\\
&=&3 l_D^2[\varphi'\chi'- a^2 V'\chi - 2a^2 V( (d-2)\psi +n\xi
-\nabla^2_{\bar{y}}G] ~~~~(a=b),
\label{aa}
\end{eqnarray}
\begin{eqnarray}                    
&G''&+ [(d-1){\cal H} + n{\cal F}] G'- \nabla^2_{\bar{x}}G
-\frac{a^2}{b^2}\nabla^2_{\bar{y}}G 
\nonumber\\
&-& \frac{a}{b} C' 
-\frac{a}{b} [d{\cal H} + (n-1) {\cal F}] C + 2(\psi- \xi) \frac{a^2}{b^2}=0
~~~(a\neq b),
\label{anb}
\end{eqnarray}
The linear system of differential equations with time dependent coefficients
formed by the three constraints (\ref{0i}), (\ref{0a}) and (\ref{aj})and by the
equations (\ref{00}), (\ref{ii}), (\ref{aa}), (\ref{anb}) determines the
classical space-time evolution of the five fluctuations $\psi$, $\xi$, $G$,
$C$ and $\chi$. In order to simplify the system we also write the perturbed
equation of motion for the scalar field which can be obtained from the
combination of the other equations:
\begin{eqnarray}
&-& \nabla^2 \chi +\chi''+[(d-1){\cal H} +n{\cal F}] \chi' 
- 2 \varphi'[(d-1) \psi' + n\xi']
\nonumber\\
&-& V''a^2 \chi + 2 a^2 V ( (d-2)\psi + n\xi
-\nabla^2_{\bar{y}}G)=0~~~(\chi).
\label{dilpert}
\end{eqnarray}
Subtracting eq. (\ref{ii}) from eq. (\ref{00}) and eq. (\ref{aa}) from
(\ref{00}) we get respectively:
\begin{eqnarray}
\Box\lambda + 3[(d-1){\cal H} + n{\cal F}]\lambda'
- (2 (\nabla^2_{\bar{y}}G)'
- \frac{a}{b}(\nabla^2_{\bar{y}}C)) [{\cal H} + \frac{n}{d-1}{\cal
F}]                          
\nonumber\\
=\frac{6l_D^2}{d-1}( a^2 V'\chi + 2 a^2 V ((d-2)\psi + n\xi-
\nabla^2_{\bar{y}}G), 
\label{lambda}
\end{eqnarray}
(where $\lambda = \psi +\frac{n}{d-1}\xi$,
$\Box=(\partial/\partial\eta)^2 -\nabla_{\bar{x}}^2 -
\frac{a^2}{b^2}\nabla^2_{\bar{y}}$) and
\begin{eqnarray}
&-&(n-1)\{\Box\xi +\xi'\left[ \frac{{\cal H}}{n-1}\left(3dn-d-n
+1\right)+3 n {\cal F}\right]\}
\nonumber\\
&+&d\{\Box\psi+\psi'\left[3(d-1){\cal H}+
\frac{{\cal F}}{d}\left( 2(d-1)(n-1)+nd\right) \right]\}
\nonumber\\
&-&(2 (\nabla^2_{\bar{y}}G)'
- \frac{a}{b}(\nabla^2_{\bar{y}}C)) [d{\cal H} + (n-1){\cal F}]     
\nonumber\\
&-&\frac{6l_D^2}{d-1}[ a^2 V'\chi + 2 a^2 V ((d-2)\psi + n\xi-
\nabla^2_{\bar{y}}G)] =0.
\label{anisotper}
\end{eqnarray}
Combining now the constraints (\ref{aj}) with eq. (\ref{anb}) we get an useful
expression which allows to eliminate $C$ from the other equations
\begin{equation}
C=\frac{1}{2({\cal H} -{\cal F})}[G''+ G'((d-1){\cal H} + n{\cal F})  
-\nabla^2_{\bar{x}}G -\frac{a^2}{b^2}\nabla^2_{\bar{y}}G]~~~~.
\label{c}
\end{equation}
Using (\ref{c}) in (\ref{lambda}) and (\ref{anisotper}) together with the
background equations in the form (\ref{third}) it is possible to show, by
linearly combining the obtained relations, that 
the evolution equations for the longitudinal fluctuations can be 
written in the gauge (\ref{pert}) as
\begin{eqnarray}
v''- \frac{z_1''}{z_1} v - \nabla^2_{\bar{x}}v -
\frac{a^2}{b^2}\nabla^2_{\bar{y}}v=0,
\label{eqvw1}
\\
w''- \frac{z_2''}{z_2} w - \nabla^2_{\bar{x}}w -
\frac{a^2}{b^2}\nabla^2_{\bar{y}}w=0,
\label{eqvw2}
\end{eqnarray} 
where
\begin{eqnarray}
v=z_2\chi + z_1 \lambda,~~~~
w= \frac{z_1}{l_D}\sqrt{\frac{n(n+d-1)}{6(d-1)}}(\frac{\cal H}{\varphi'}
 \xi - \frac{\cal F}{\varphi'} \psi)
\label{canonical}
\end{eqnarray}
and
\begin{equation}
z_1=\frac{a^{\frac{d-1}{2}}b^{\frac{n}{2}}\varphi' }{{\cal
H}+\frac{n}{d-1}{\cal F}},~~~ 
,~~~z_2=a^{\frac{d-1}{2}}b^{\frac{n}{2}}\equiv [\frac{-g}{a^2}]^{\frac{1}{4}}.
\end{equation}
Since $\phi$, $\psi$ and $\xi$ coincide, in the gauge
(\ref{gauge}), with the corresponding gauge-invariant quantities listed in eq.
(\ref{bardeen}) we can write that
\begin{eqnarray}
{\cal V}=z_2X + z_1 \Lambda,~~~~\Lambda= \Psi
+\frac{n}{d-1}\Xi,~~~~
{\cal W}= \frac{z_1}{l_D} \sqrt{\frac{n(n+d-1)}{6(d-1)}}
 (\frac{\cal H}{\varphi'} \Xi - \frac{\cal F}{\varphi'} \Psi).
\label{gi}
\end{eqnarray}
Notice that in the absence of internal dimensions ${\cal W}$ is zero and
${\cal V}$ coincides with the scalar normal mode of oscillation which
diagonalizes the the action (\ref{4}) perturbed to second order in the amplitude
of the fluctuations \cite{9}.
By solving the equations for ${\cal V}$ and ${\cal W}$ and by using their
definition in terms of the gauge-invariant fluctuations it is possible to
obtain the time evolution of all the quantities listed in (\ref{bardeen}) (the
explicit solution for the longitudinal fluctuations will be discussed in Sec.
5).

The evolution equations for the gauge-invariant tensor modes 
(only coupled to the scalar curvature and not to the sources of the
background) can be obtained without any specific gauge choice.
The form of the perturbed metric will be in this case :
\begin{equation}
\delta g^{(T)}_{\mu\nu}= \left(\matrix{0&0&0&\cr
0&-a^2 h_{ij}&0&\cr
0&0& -b^2 H_{ab}&\cr}\right)
\end{equation}
with $\nabla_i h^{ij}= h_i^i =0$ and  $\nabla_a H^{ab}= H_a^a =0$ (since
$H_a^b$ and $h_i^j$ are pure tensor modes in each space
they are also automatically gauge-invariant with respect to gauge
transformations preserving the tensor nature of the fluctuations in the
external and internal manifold).  
The evolution equations can be easily written by perturbing
the Ricci tensor. All the components of the Einstein equations (\ref{pertein})
are zero but ($i,j$) and ($a,b$):
\begin{eqnarray}
h''+ [(d-1) {\cal H} +n {\cal F}] h' -\nabla^2_{\bar{x}}h
-\frac{a^2}{b^2}\nabla^2_{\bar{y}}h =0,
\nonumber\\
H''+ [(d-1) {\cal H} +n {\cal F}]H' -\nabla^2_{\bar{x}}H
-\frac{a^2}{b^2}\nabla^2_{\bar{y}}H =0,
\label{h}
\end{eqnarray}
where $h_{i}^{j}\equiv h(\eta, \overline{x}, \overline{y}) e_i^j$ 
and $ H_a^b \equiv H( \eta, \overline{x}, \overline{y}) E_a^b$ ( $e_i^j$ and
$E_a^b$ are, respectively the external and internal polarization tensors).  
As we already reminded the gravity waves polarized along the internal
dimensions will not be able to excite a detector 
of tensor waves but will be seen in the ($d+1$)-dimensional ``external" world as
density fluctuations and this means, from the mathematical point of view, 
that $h_i^j$ are scalar eigenstates of the internal Laplace
operator ($\nabla^2_{\bar{y}}$) in the same 
way as  $H_a^b$ are scalar eigenstates of the external Laplace operator 
($\nabla^2_{\bar{x}}$) (of course $h_i^j$ and $H_a^b$ are also solution of the 
tensor Helmotz equation, respectively, in the external and internal manifold).
By defining $\mu =(1/24 l_D) a^{\frac{d-1}{2}}b^{\frac{n}{2}} h$ and
 ${\cal M}=(1/24 l_D) a^{\frac{d-1}{2}}b^{\frac{n}{2}} H$, 
equations (\ref{h}) can be easily rewritten as :
\begin{eqnarray}
\mu'' -\nabla^2_{\bar{x}}\mu -\nabla^2_{\bar{y}}\mu -
\frac{z_2''}{z_2}\mu=0,~~~~~~
{\cal M}'' -\nabla^2_{\bar{x}}{\cal M} -\nabla^2_{\bar{y}}{\cal
M} - \frac{z_2''}{z_2}{\cal M}=0
\label{canonical2}
\end{eqnarray}
We notice that in the absence of the internal dimensions $\mu$ coincide with 
the amplitude of the tensor normal modes of oscillation of the action (\ref{4})
\cite{9}.
The solution of the coupled systems of differential equations which describe
the evolution of the scalar and tensor inhomogeneities was previously
investigated \cite{11,14} well outside  the horizon (where the internal
and external Laplace operators are subleading)
 in the case of the Kaluza-Klein 
``vacuum" solutions (\ref{sol1}) and in the case of the dilaton-driven solutions
(\ref{sol2}).  
Provided we neglect the internal and external Laplacians
 it can be actually shown quite easily
that the  solution to eq. (\ref{canonical2}) is :
\begin{equation}
\mu(\eta) = a^{\frac{d-1}{2}} b^{\frac{n}{2}}
 ( A_1 + B_1 \int \frac{ a d\eta}{a^d b^n}), ~~~~
 {\cal M}(\eta) = a^{\frac{d-1}{2}} b^{\frac{n}{2}}
 ( A_2 + B_2 \int \frac{ a d\eta}{a^d b^n}), ~~~~
\label{s1}
\end{equation}
where $A_i$, $B_{i}$ are the integration constants.
Since from (\ref{sol1}) and (\ref{sol2})
 $a^{\frac{d-1}{2}}b^{\frac{n}{2}}\equiv \sqrt{|\eta|}$ for both the
 backgrounds in an arbitrary number of dimensions, we will have (from eq.
 (\ref{s1})) that outside the horizon the gravity wave amplitude diverges
at most logarithmically :
\begin{equation}
\mu\simeq \sqrt{|\eta|}\left(A_1 + B_1 \ln{|\eta|}\right),~~~~~
{\cal M}\simeq \sqrt{|\eta|}\left(A_2 + B_2 \ln{|\eta|}\right).
\label{log}
\end{equation}
In the case of the scalar
inhomogeneities it can be shown instead (for example from eq. (\ref{lambda})
neglecting the Laplacians and for $V=0$)
that the longitudinal fluctuations diverge like a power outside the horizon for
both the background solutions (\ref{sol1})-(\ref{sol2}) and typically we will
have: 
\begin{equation}
\Lambda \simeq~~c_1 +
\frac{c_2}{\eta^2} 
\end{equation}
(a similar behaviour can be deduced also for $\Psi$ and $\Xi$).
If growing solutions are present it is in general not possible to keep the
amplitude of the fluctuations small all the time and at some point the
perturbative approach will unavoidably break down leading to the so called
``growing-mode problem" (which will be addressed in Sec. 5).
In order to reliably compute the spectrum of the scalar and tensor fluctuations
it seems important to consider explicitly the contribution of the internal
and external Laplacians and since this could be quite difficult for a 
generic multidimensional background we will limit our attentions to the 
models described in eq. (\ref{toy1}) and (\ref{toy2}) which will be analyzed in
the following two Sections.

\renewcommand{\theequation}{4.\arabic{equation}}
\setcounter{equation}{0}
\section{Graviton production from extra spatial dimensions}

The spectral energy density of the cosmic gravitons produced thanks to the time
evolution of an homogeneous and isotropic cosmological model is an important
source of dynamical informations \cite{1G,1bG}
 since the slopes of the spectra versus the
frequency offer a snapshot of the early history of the hubble parameter 
\cite{2G,3G}. Moreover the graviton spectra can indirectly constrain 
the homogeneous and isotropic inflationary models \cite{4G,5G}.
 If we relax the assumption
of isotropy of the background manifold it is unclear how to perform the
calculation of the spectral amplitudes which will have to be eventually
compared to the phenomenological bounds 
(see \cite{5G} and \cite{6G} for two reviews concerning the stochastic
gravity-waves backgrounds and their detectability). 
Our purpose is to calculate the
graviton spectra produced in the two oversimplified models of dimensional
decoupling presented in Sec. 2 in order to get the feeling of what could happen
in a realistic situations. We will see that thanks to the coupling among
the scalar and the tensor modes the gravity wave evolution equation will get a
``massive " contribution which might also be  relevant in the case of more
motivated  background geometries. 
In terms of the eigenstates of the Laplace operators
\begin{eqnarray}
\nabla^2_{\bar{x}}h_i^j(k,q) = - k^2 h_i^j(k,q)
\nonumber\\
\nabla^2_{\bar{y}}h_i^j(k,q) = - q^2 h_i^j(k,q),
\end{eqnarray}
eq. (\ref{canonical2}) will be
\begin{equation}
\mu''+ [k^2 + q^2 \frac{a^2}{b^2} -
\frac{z_2''}{z_2}]\mu =0~~~.
\label{wave}
\end{equation}
We consider first of all the model (\ref{toy1}). For $\eta\leq -\eta_c$  
eq. (\ref{wave}) becomes :
\begin{equation}
\mu''+[k^2 + \frac{1}{4\eta^2}+ \frac{q^2 \eta_c^2}{\eta^2}]\mu=0,
\label{gw1}
\end{equation}
whereas for $\eta\geq -\eta_c$ the same equation will be
\begin{equation}
\mu''+[k^2 + \frac{q^2}{\eta_c^2}(\eta +2\eta_c)^2]\mu=0.
\label{gw2}
\end{equation}
For $\eta<-\eta_c$ an exact solution of eq. (\ref{gw1}) can be written in term 
of the Hankel functions:
\begin{equation}
\mu(k\eta,q) = \frac{1}{\sqrt{k}} \sqrt{k\eta} H_{\nu}^{(2)}(k\eta),~~~~
\nu= i (q\eta_c)
\label{bessel}
\end{equation}
(we have chosen the positive frequency mode which corresponds [for
$\eta\rightarrow-\infty$] to the Bunch-Davies vacuum).
In the absence of the internal Laplacians ( or, equivalently, if we would
keep only the lowest mode of the internal excitations, $q=0$) instead of
(\ref{bessel}) we would get a completely different solution:
\begin{equation}
\mu(k\eta) = \frac{1}{\sqrt{k}}\sqrt{k\eta} H_0^{(2)}(k\eta)
\label{0bess}
\end{equation}
whose limit for small arguments holding when the given mode is well outside 
the horizon ($k\eta<<1$) gives :
\begin{equation}
\mu(k\eta)= \sqrt{\eta} ( 1 -\frac{2i}{\pi} \ln{k\eta})
\label{log2}
\end{equation}
which is clearly consistent with the evolution of the gravity waves outside the
horizon in an arbitrary number of dimensions (and for generic initial
conditions)
obtained in (\ref{log})
 For $\eta>-\eta_c$ the
solution of eq. (\ref{gw2}) can be written in terms of parabolic cylinder
functions (defining a new variable $z= \sqrt{2q/\eta_c}(\eta+
2\eta_c)$  eq. (\ref{gw1}) becomes exactly the parabolic cylinder equation
expressed in its standard form \cite{23,24}):
\begin{equation}
\mu(k\eta, q) = (\frac{2q}{\eta_c})^{\frac{1}{4}}( c_{-} E(a, z) + c_{+}
E^{\ast} (a, z))
\label{parabolic}
\end{equation}
($a = -k^2 \eta_c / 2 q$ and $E(a,z)$, $E^{\ast}(a,z)$
 are complex conjugated solutions of the parabolic
cylinder equation \cite{23,24}). If $a > z^2/ 4$  namely if $ k^2 > q^2 ((\eta
+2\eta_c) /\eta_c)^2$ we have that the solution (\ref{parabolic}) becomes :
\begin{equation}
\mu(k\eta, q) \rightarrow \frac{1}{\sqrt{k}} (c_{-} e^{-ik(\eta+2\eta_c)}+
c_{+} e^{ik(\eta+2\eta_c)}).
\label{planew}
\end{equation}
In the opposite limit ($k^2< q^2 (\frac{\eta+2\eta_c}{\eta_c})^2$) solution
(\ref{parabolic}) becomes instead
\begin{equation}
\mu(k\eta, q) \rightarrow \sqrt{\frac{\eta_c}{q(\eta+2\eta_c)}}( c_{-} 
e ^{\frac{i}{2}q\frac{(\eta + 2\eta_c)^2}{\eta_c}} +c_{+}  
e ^{-\frac{i}{2}q\frac{(\eta + 2\eta_c)^2}{\eta_c}})
\label{mass}
\end{equation}
(as can be directly obtained by solving eq. (\ref{gw1}) 
for a negligible $k^2$). The  last solution is identical to the
evolution equation of a minimally coupled scalar fields in the
 radiation dominated era with mass $m\sim q$ so that the effect of the internal
Laplacians on the evolution of an externally polarized gravity wave 
evolving during the radiation dominated era can be described with an effective
mass term whose magnitude depends on the magnitude of the excitations 
belonging to the internal space.
In the Schroedinger-like equation (\ref{wave})-(\ref{gw1}) the mass term 
modifies the potential barrier whose maximum is now $1/{4\eta_c}^2  + q^2$. 
This effective potential barrier leads to wave amplification \cite{1G,8G}, 
or, equivalently, to particle production \cite{19}.
Actually the positive frequency modes (for
$\eta\rightarrow -\infty$) in eq. (\ref{bessel}) will be in general a linear
combination  of modes which are of positive or negative frequency with respect
to the vacuum to the right ($\eta\rightarrow +\infty$).
The coefficients of the Bogoliubov transformation ($c_{\pm}$, $|c_+|^2 -
|c_-|^2=1$) connecting the left and right vacuum and fixed by  matching,
in $\eta=-\eta_c$, each solution and its first derivative will determine the
spectral density of the produced gravitons. 
We can now compute the amplification coefficient $c_{-}$ in the sudden
approximation \cite{9G}, namely for $(k\eta_c)^2 <  1 + (q\eta_c)^2$ (for 
$(k\eta_c)^2 > 1 + (q\eta_c)^2$ there is no wave amplification 
 and the Bogoliubov
coefficient $c_{-}$ is exponentially suppressed ).
 We will have in general a two branches amplification coefficient 
depending if the mode $k$ is ``non-relativistic"   
($ k^2 < q^2 (\frac{\eta + 
2\eta_c}{\eta_c})^2 $) or ``ultrarelativistic" ($ k^2 > q^2 (\frac{\eta + 
2\eta_c}{\eta_c})^2 $).
So matching the solutions (\ref{bessel}) and (\ref{planew})  in $\eta=-\eta_c$
 we obtain:
\begin{equation}
e^{ik\eta_c} c_{-} \simeq \frac{1}{2\sqrt{2}}[(\frac{1}{2}-\nu) 
\frac{\Gamma(\nu)}{\pi} ( \frac{k\eta_{c}}{2})^{-\nu-\frac{1}{2}} - i 
(\frac{k\eta_c}{2})^{\nu -\frac{1}{2}} (\nu +\frac{1}{2}) \frac{1}
{\nu\Gamma(\nu)}], 
\label{bog1}
\end{equation}
for $ k^2 > q^2 \frac{(\eta +2\eta_c)^2}{\eta_c^2}$, and
\begin{equation}
e^{i\frac{q\eta_c}{2}}c_{-} \simeq  \frac{\Gamma(\nu)}{2\pi}
[\sqrt{q\eta_{c}} + 
(1-\nu)(q\eta_{c})^{-\frac{1}{2}}](\frac{k\eta_{c}}{2})^{-\nu}
\label{bog2}
\end{equation}
for $k^2 < q^2 \frac{q^2 (\eta + 2\eta_c)^2}{\eta_c^2}$. 
The typical amplitude of gravity waves over scales $k^{-1}$  is $\delta
h(k,q,\eta)
= l_D k^{d/2} q^{n/2}|h(k,q,\eta)|$ (where $l_D= M_P^{\frac{n+d-1}{2}}$) and
can be easily computed using the definition of $\mu$ in terms of $h$ (from
equations (\ref{h}) and (\ref{canonical2}))
\begin{equation}
\delta h(\omega) = (\frac{\omega}{\omega_c}) \frac{H_c}{M_P}
z_{dec}^{-\frac{1}{4}} (\frac{\omega_q}{\omega_c}) \sqrt{\frac{\omega_0}{M_P}}
 |c_-|
\label{pw}
\end{equation}
($\omega= k/a$, $\omega_q= q/a$, $\omega_c= H_c a_c /a $ is the maximal
amplified frequency and $\omega_0 = 3.1 \times 10^{-18} ~ h_{100}~Hz$ is the 
present value of the Hubble parameter; we used that, in our case, $a(\eta_c)
\simeq b(\eta_c)\simeq 1$).
Keeping only the leading terms for $k\eta_c<1$ in eq. (\ref{bog1}) and
(\ref{bog2}) the power spectrum (\ref{pw}) is:
\begin{eqnarray}
|\delta h(\omega)| &\simeq& z_eq^{-\frac{1}{4}}
 (\frac{\omega}{\omega_c})^{\frac{1}{2}}
(\frac{H_c}{M_P})^{\frac{1}{2}}
(\frac{\omega_0}{M_P})^{\frac{1}{2}}
(\sinh{\pi(\frac{\omega_q}{\omega_c})})^{-\frac{1}{2}},~~~~\omega>\omega_q
\nonumber\\
|\delta h(\omega)| &\simeq& z_eq^{-\frac{1}{4}}
 (\frac{\omega}{\omega_c})
(\frac{H_c}{M_P})
(\frac{\omega_0}{M_P})^{\frac{1}{2}}
(\sinh{\pi(\frac{\omega_q}{\omega_c})})^{-\frac{1}{2}},~~~~\omega<\omega_q
\label{spec1}
\end{eqnarray}
(where we used that $\Gamma(iq\eta_c) \Gamma(-iq\eta_c) = \pi/ (q\eta_c
\sinh{q\eta_c})$ \cite{23}). Since we assumed that the radiation starts
dominating suddenly after $-\eta_c$, we can estimate that $\omega_c = 10^{11}
~~\sqrt{H_c/M_P}~~Hz$, assuming that the evolution is adiabatic ( if the
evolution is not adiabatic and entropy is produced at some stage
 this result could be slightly modified but for
our illustrative purposes it is not crucial \cite{6G} [see however \cite{10G}
for a more quantitative analysis, in a more specific four dimensional model]). 
In order to compare  the power spectrum with the
phenomenological bounds which could constrain the parameter space of our
 naive model it is useful to work with the fraction of critical
density stored in the gravity wave background per logarithmic
interval of $\omega$:
\begin{equation}
\Omega_{GW}(\omega)=\frac{1}{\rho_c}\frac{d \rho_{GW}}{d\ln{\omega}} = 
z_{dec}^{-1} \left(\frac{\omega_q}{\omega_c}\right)
\left(\frac{H_c}{M_P}\right)^3 |c_{-}|^2 \simeq 
\left(\frac{\omega}{\omega_0}\right)^2 |\delta h(\omega)|^2,~~~\rho_c\sim
l_{D}^2 H_c^2 (a_c/a)^4.
\end{equation}
Using again the explicit expression for the Bogoliubov coefficients (\ref{bog1})
-(\ref{bog2}) in the two different regimes we get:
\begin{eqnarray}
\Omega_{GW}(\omega) &\simeq& z_{dec}^{-1} \left(\frac{\omega}{\omega_c}\right)^3
\left(\frac{H_c}{M_P}\right)^3 (\sinh{\pi(\frac{\omega_q}{\omega_c})})^{-1}
,~~~~~\omega>\omega_q
\nonumber\\
\Omega_{GW}(\omega) &\simeq& z_{dec}^{-1} \left(\frac{\omega}{\omega_c}\right)^4
\left(\frac{\omega_q}{\omega_c}\right)^{-1}
\left(\frac{H_c}{M_P}\right)^3 (\sinh{\pi(\frac{\omega_q}{\omega_c})})^{-1}
,~~~~~\omega<\omega_q
\label{endens}
\end{eqnarray}
($H_c/M_P$ measures how far from the Planck scale the compactification 
occurs and $\omega_q/\omega_c$ estimates the typical frequency of the internal
excitations $\omega_q$ evaluated at the beginning of the radiation epoch
($\eta=-\eta_c$) with respect to the maximal amplified frequency $\omega_c$;
notice also that since $\omega_c$ is the maximal amplified frequency 
$|\omega_q/\omega_c|<1$).
While the amplitude of the spectra are characterized by the two
dimension-less quantities $x=\log_{10}{(\omega_{q}/\omega_c)}$ and
 $y= \log_{10}{(H_c/M_P)}$, the spectral slope is instead fixed by the 
background evolution and can be also more difficult to estimate in a different
model of dimensional decoupling. In our case  the parameter space 
 can be constrained by the observations and since the spectrum is
increasing in frequency we would expect that the bounds coming from the large 
scales like the $COBE$ bound \cite{11G}
 ($\Omega_{GW}(\omega) < 7.1 \times 10^{-11}$ for
$\omega_0<\omega<30\omega_0$) and the pulsar bound \cite{12G}
 ($\Omega_{GW}<10^{-8}$ at
$\omega \sim 10^{-8}~ Hz$) will be less constraining than the bounds arising
from nucleosynthesis \cite{13G} ($\int d\ln{\omega} 
\Omega_{GW}(\omega, \eta_0)< 0.2
~\Omega_{\gamma}(\eta_0) \sim 10^{-5}$, [$\Omega_{\gamma}(\eta)$ is the fraction
of critical energy density present in form of radiation, at a given observation
time $\eta$]) or from the critical energy density  ($\Omega(\omega)<1$ for
all the frequencies). In particular from (\ref{endens}) we can find that the
COBE bound is satisfied either if $y\laq 0.6~x + 35 - 2 \log_{10}{h_{100}}$
(provided $\omega_{COBE}\simeq 10 ~\omega_0< \omega_{q}$)  
or if $ y \laq 0.6~x +8 -3\log_{10}{h_{100}}$ (for
$\omega_{COBE}>\omega_{q}$); the pulsar bounds are satisfied either
if $y \laq 0.3~x +35$ (for $\omega_{q}<\omega_{p}$) or if $y\laq 2~x +72$
($\omega_{q}>\omega_{p}$).
The $COBE$ and the pulsar bounds are less constraining,
while the critical density and the nucleosynthesis bound combined together
give $y\laq -0.6~x -0.6$ (for $\omega< \omega_{q}$) and $y\laq 0.3~x -0.6$
(for $\omega> \omega_{q}$) which is compatible with $H_c< 10^{-1} M_P$
and $\omega_q < \omega_{c}$.

A similar analysis can be performed in the case of the model (\ref{toy2}). 
The evolution equation (\ref{wave}) is then (for $\eta<-\eta_c$):
\begin{equation}
\mu''+ \left[ -\frac{1}{4} +\frac{a}{z} +\frac{1}{4z^2}\right]\mu=0,
~~~~z=2ik(\eta-\eta_c),~~~a=i\frac{q^2\eta_c}{k}
\end{equation}
which is formally equivalent to the radial Schroedinger equation for the
problem of the Coulomb diffusion and which can be easily solved in terms 
of Confluent Hypergeometric functions 
\begin{equation}
\mu(k\eta,q)= e^{-ik(\eta-\eta_c)} \sqrt{2k(\eta-\eta_c)}U(\frac{1}{2}-
i\frac{q^2\eta_c}{k},~1,~2ik(\eta-\eta_c)),~~~~\eta<\eta_c
\label{kum}
\end{equation}
($U$ is the Kummer function defined with the conventions of \cite{23}; for 
$\eta\rightarrow -\infty$ the solution
 behaves like a positive frequency mode , but
does not define, asimptotically, a Bunch-Davies adiabatic vacuum). As in the 
previous example we have to match the solution (\ref{kum}) valid for
$\eta<-\eta_c$with the solutions (\ref{planew}) and (\ref{mass}) valid for
$\eta>-\eta_c$. The result of this procedure will give the Bogoliubov
coefficients describing the mixing positive and negative frequency modes:
\begin{eqnarray}
c_{-}&=&e^{-2i k\eta_c}[\frac{i}{4\sqrt{2k\eta_c}} U(\frac{1}{2}-
i\frac{(q\eta_c)^2}{k\eta_{c}},~1,~-4ik\eta_c) 
\nonumber\\
&+& \sqrt{2k\eta_c} U(\frac{1}{2}-
i\frac{(q\eta_c)^2}{k\eta_{c}},~1,~-4ik\eta_c)]
,~~~k^2> q^2 (\frac{\eta + 2\eta_c}{\eta_c})^2
\nonumber\\
c_{-}&=&e^{-i \frac{q\eta_c}{2}}[(\sqrt{2q\eta_c} -\sqrt{2k\eta_c})
 U(\frac{1}{2}-i\frac{(q\eta_c)^2}{k\eta_{c}},~1,~-4ik\eta_c) 
\nonumber\\
&+&\sqrt{\frac{k}{q}} \sqrt{2k\eta_c} U(\frac{1}{2}-
i\frac{(q\eta_c)^2}{k\eta_{c}},~1,~-4ik\eta_c)]
,~~~k^2< q^2 (\frac{\eta + 2\eta_c}{\eta_c})^2
\end{eqnarray}
Using the small argument limit of the Kummer functions we can compute the
normalized spectral amplitude:
\begin{eqnarray}
\delta h(\omega)&\simeq& 
z_{eq}^{-\frac{1}{4}} \left(\frac{\omega_q}{\omega_c}\right)^3
\left(\frac{H_c}{M_P}\right)^{\frac{7}{2}} \left(\frac{\omega}{\omega_c}\right)
^{\frac{1}{2}},~~~\omega>\omega_q
\nonumber\\
\delta h(\omega)&\simeq& 
z_{eq}^{-\frac{1}{4}} (\sinh{\pi\frac{\omega_q}{\omega_c}})^{-\frac{1}{2}}
\left(\frac{H_c}{M_P}\right)^{\frac{7}{2}}\left(\frac{\omega_0}{M_P}\right)
^{\frac{1}{2}}
 \left(\frac{\omega}{\omega_c}\right)
,~~~\omega<\omega_q
\label{spec2}
\end{eqnarray}
and the spectral energy density distribution in critical units:
\begin{eqnarray}
\Omega_{GW} &\simeq& z_{dec}^{-1}
\left(\frac{\omega_q}{\omega_c}\right)^6 
\left(\frac{H_c}{M_P}\right)^8 
\left(\frac{\omega}{\omega_c}\right)^3 \ln{(\frac{\omega}{\omega_c})}
~~~~~\omega>\omega_q
\nonumber\\
\Omega_{GW} &\simeq& z_{dec}^{-1} \left(\frac{\omega_q}{\omega_c}\right)^6
\left(\frac{H_c}{M_P}\right)^8 (\sinh{\pi(\frac{\omega_q}{\omega_c})})^{-1}
\left(\frac{\omega}{\omega_c}\right)^{4}
~~~~~\omega<\omega_q
\end{eqnarray}
For $\omega>\omega_q$ the slopes of the spectral energy density distribution 
(and of the related spectral amplitudes) agree with the result previously 
obtained for the spectra of gravity waves produced during a dilaton driven 
phase \cite{18b}, neglecting the internal Laplace-Beltrami operators. 
The amplitudes are instead different due to the presence of 
$\omega_q$. Our partial conclusion is  that to neglect the internal 
Laplacians is a good approximation for the slopes of the spectra  
(for $\omega>\omega_q$) but not for their amplitude. 
For $\omega<\omega_q$ both the slopes and the amplitudes of the spectra
are affected by the presence of the internal Laplacians which cannot be
overlooked.
Since the spectra are increasing in frequency we will keep the most stringent
bound which comes from nucleosynthesis and which gives, if applied separately
in each of the two branches, $y\laq -0.7~x -0.2$ (for $\omega>\omega_q$),
and $y\laq -1.1~x-0.25$ (for $\omega<\omega_q$).
 
The obtained spectral amplitudes (\ref{spec1})-(\ref{spec2}) are quite 
different since they are produced by two different background geometries, 
but the spectral slopes are exactly equal in spite of the differences in the 
solutions (\ref{sol1})-(\ref{sol2}). More specifically we obtained 
 ``violet" type of spectra ($\delta h\sim \omega/\omega_c$ for 
$\omega>\omega_q$  and  $\delta h\sim (\omega/\omega_c)^{1/2}$ 
for $\omega<\omega_q$) which are a common feature of the contracting 
backgrounds also in the isotropic case \cite{7vG}.
This apparent puzzle is due to the fact that $a^{\frac{d-1}{2}} 
b^{\frac{n}{2}} \sim \sqrt{|\eta/\eta_c|}$ 
for (\ref{sol1}) and (\ref{sol2}) in arbitrary number of internal and 
external dimensions.

We would like finally to point out that the graviton production due to the
transition from the radiation dominated stage to the matter dominated stage
should also be included. This further amplification will modify the low
frequency tail  of the spectrum ($10^{-18} ~Hz< \omega< 10^{-16}~Hz$). 
The qualitative
aspects of our analysis  show that the presence of the internal gradients 
introduces in the spectral amplitude
 a new parameter related to the frequency of the internal
excitation which can be interestingly constrained by the bounds 
usually analyzed in the context of the stochastic gravity-wave backgrounds.

\renewcommand{\theequation}{5.\arabic{equation}}
\setcounter{equation}{0}
\section{Growing solutions for the scalar modes}

The tensor inhomogeneities can be treated perturbatively keeping track of
the internal Laplacians also because they evolve  logarithmically 
outside the horizon.The situation changes in the case of the scalar 
inhomogeneities because, as we explicitly pointed out at  the end of Sec. 2,
the growing solution increases much faster than a logarithm and
a perturbative treatment become inappropriate after some time. The hope would
be in this case that the growing mode appearing in the multidimensional case
could be consistently gauged down as happens for the dilaton driven solutions
in the $(3+1)$-dimensional case\cite{14}.
In this section we do not want to address 
specifically the problem of the growing modes in an anisotropic
manifold but we want to show how the
 problem can be consistently formulated in the presence of the
internal Laplacians 
and for this purpose we will study 
the $10$-dimensional dilaton-driven solutions (\ref{sol2}).
Using the all set of equations (\ref{00})-(\ref{anb}) and equations
(\ref{eqvw1}), (\ref{eqvw2}), (\ref{canonical}) the Fourier modes of the
longitudinal fluctuations $\Psi$, $\Lambda$ can be expressed in terms of the 
Fourier modes of ${\cal V}$ and ${\cal W}$
\begin{eqnarray}
(k^2 + \frac{a^2}{b^2} q^2)\Psi(k,q,\eta) &=& \frac{ n(n+d-1) {\cal H}{\cal
F}\varphi'}{[(d-1){\cal H}+ n{\cal
F}]^2}\left[\frac{6l_D^2(d-1)}{n(n+d-1)}\right]^{1/2}
\left(\frac{{\cal W}(k,q,\eta)}{z_1}\right)'-
\nonumber\\
&-& \frac{3l_D^2\varphi' {\cal H}}{[(d-1){\cal H} +n {\cal F}]}
\left(\frac{{\cal V}(k,q,\eta)}{a^{\frac{d-1}{2}}b^{\frac{n}{2}}}\right)'
\nonumber\\
&-&\frac{n\varphi'}{(d-1){\cal H} + n{\cal F}}
\left[\frac{6l_D^2(d-1)}{n(n+d-1)}\right]^{1/2}
\left(\frac{{\cal W}(k,q,\eta)}{z_1}\right)~ ,
\label{psivw}
\\
(k^2+\frac{a^2}{b^2}q^2) \Lambda(k,q,\eta) &=&\frac{n {\cal F}}{d-1} \frac{
[(n+d-1)\varphi']}{[(d-1){\cal H} + n{\cal
F}]}\left[\frac{6 l_D^2(d-1)}{n(n+d-1)}\right]^{1/2}
\left(\frac{{\cal W}(k,q,\eta)}{z_1}\right)'
\nonumber\\ 
&-& \frac{3 l_D^2\varphi'}{(d-1)}\left(\frac{{\cal V}}{a^{\frac{d-1}{2}}
b^{\frac{n}{2}}}\right)'
\label{lambdavw}
\end{eqnarray}
where $q$, $k$ have to be considered both scalar eigenvalues (while in the
previous section $k$ was labeling the eigenvalues of the tensor Helmotz
equation). If $k^2 > q^2 a^2/b^2$ the evolution equation 
for ${\cal V}(k,q,\eta)$
and ${\cal W}(k,q,\eta)$ will be (from (\ref{eqvw1}), (\ref{eqvw2}) through 
(\ref{canonical}))
\begin{equation} 
{\cal V}'' + \left[k^2 + \frac{1}{4(\eta_c-\eta)^2}\right]{\cal V}=0,~~~
{\cal W}'' + \left[k^2 + \frac{1}{4(\eta_c-\eta)^2}\right]{\cal W}=0
\end{equation}
whose solution is exactly identical to (\ref{log2}), provided 
we choose, for $\eta\rightarrow-\infty$, the Bunch-Davies vacuum
as initial condition. 
In the limit
$k^2 < q^2 a^2/b^2$ eq. (\ref{eqvw1})-(\ref{eqvw2}) and (\ref{canonical}) 
will give instead
\begin{equation}
{\cal V}'' + \left[\frac{\eta_c q^2}{(\eta_c-\eta)}
 + \frac{1}{4(\eta_c-\eta)^2}\right]{\cal V}=0,~~~
{\cal W}'' + \left[\frac{\eta_c q^2}{(\eta_c-\eta)} 
+ \frac{1}{4(\eta_c-\eta)^2}\right]{\cal W}=0
\label{I}
\end{equation}
with solution  \cite{23}
\begin{equation}
{\cal V}(k,q,\eta)= 
\sqrt{\eta_c-\eta} H_{0}^{(2)}(q\sqrt{\eta_c-\eta}),~~
{\cal W}(k,q,\eta)= 
\sqrt{\eta_c-\eta} H_{0}^{(2)}(q\sqrt{\eta_c-\eta}),~~
\label{II}
\end{equation}
The typical amplitude of the longitudinal fluctuations  
over a length scale $k^{-1}$ is $|\delta \Psi(k,q,\eta)| = l_D k^{\frac{d-1}{2}}
q^{\frac{n}{2}} \Psi(k,q,\eta)$ (with $l_D= M_P^{\frac{d+n-1}{2}}$). In the
particular case of the $10$-dimensional model (\ref{toy2}) we have  
$|\delta\Psi(kq,\eta)| =l_D k^{3/2} q^3 
\Psi_{k,q}(k,q,\eta)$ and from (\ref{psivw})
with the use of (\ref{I})  we obtain for $k^2 > q^2 (a/b)^2$
\begin{equation}
|\delta\Psi|(k,q,\eta) \simeq\left(\frac{H_c}{M_P}\right)^4 
\frac{(k\eta_c)^{3/2}}{(k\eta)^2}
\left(\frac{\omega_q}{\omega_c}\right)^2
\end{equation}
(we used that $z_1 = \sqrt{|\frac{\eta}{\eta_c}|}$ and $H_{0}^{(2)}(z)\sim 
\ln{z}$).
Since $k_c\sim 1/\eta_c$ we find that $|\delta\Psi(k,q,\eta)|<1$ 
on scales $k^{-1}$ such that 
$|\eta/\eta_c|> (H_c/M_P)^2 (\omega_q/\omega_c)^{3/2}(k_c/k)^{1/4}$.
From eq. (\ref{psivw}) using (\ref{II}) we obtain, for $k^2<q^2 (a/b)^2$
\begin{equation}
|\delta\Psi|\simeq \left(\frac{H_c}{M_P}\right)\left(\frac{\omega_q}{\omega_c}
\right)\left(\frac{k}{k_c}\right)^{3/2} \left(\frac{\eta_c}{\eta}\right)
\renewcommand{\theequation}{6.\arabic{equation}}
\end{equation}
which implies that the perturbative approach  is only reliable
for conformal times \newline
$|\eta/\eta_c|> (H_c/M_P)^4(\omega_q/\omega_c)(k/k_c)^{3/2}$.

The presence of the internal gradients 
slightly changes the quantitative estimates but does not change the nature
of the growing mode problem. If this is the situation the scalar fluctuations
will become very soon critical and will then enter in a true non-perturbative
regime. This apparent contradiction among the behaviour of the tensor
inhomogeneities and the behaviour of the scalar ones might be the connected
with our perturbative technique. The fluctuations which we are discussing now
are gauge-invariant only for infinitesimal coordinate transformations while some
quantities which are invariant to all orders would be more suitable for the
study of fluctuations which are growing outside the horizon. 
It can be actually shown
\cite{14} that in the $(3+1)$ isotropic case the linearized variables
describing the scalar and tensor fluctuations in a fully covariant and
gauge-invariant approach 
\cite{8} grow only logarithmically outside the horizon.
 Unfortunately a fully covariant and gauge-invariant formalism is
only formulated in the case of a homogeneous and isotropic manifold and it
seems to be quite complicated to formulate it  for an anisotropic manifold.
Within the linearized theory discussed in this paper it would be
anyway interesting
 to understand if it is 
possible to ``gauge-down" the growing mode solutions arising in a
anisotropic background perhaps with a  suitable
 generalization of the
``off-diagonal" gauge \cite{14,22b} to the case of an anisotropic background. 

\setcounter{equation}{0}

\section{Conclusions}

We discussed the treatment of the scalar and tensor inhomogeneities in an
anisotropic background manifold undergoing dimensional decoupling.
 
We showed that it is possible to study the evolution equations of the scalar
modes by completely fixing the coordinate system with a suitable gauge choice
which reduces, in the isotropic case, to the well known conformally newtonian
gauge often employed in the analysis of the density fluctuations in the context
of the inflationary models driven by a scalar field or by perfect fluid matter.
The coupled system of second order differential equations describing the 
fluctuations of the scalar field and of a homogeneous and anisotropic 
manifold was diagonalized in terms of two scalar variables which become, 
in the absence of internal dimensions, the normal modes of oscillation
obtained by perturbing scalar-tensor action to second order in the amplitude
of the fluctuations.   
The evolution equations for the amplitude of the scalar perturbations (depending
on the internal and external coordinates) were also explicitly solved in a
particular $10$-dimensional background motivated by String Cosmology.
The scalar spectral amplitudes grow outside the horizon faster than the tensor
amplitudes and the dependence on the internal coordinates does not change
drastically this situation unless the present value of the typical frequency 
of the internal dimensions would be much smaller (by several orders of
magnitude) than the maximal amplified frequency.

The scalar and tensor modes are coupled because the tensors defined in the
internal manifold are scalar eigenstates of the Laplace-Beltrami operators 
defined in the external manifold. In order to study the connection among the
occurrence of a process of dimensional reduction and the present energy
distribution of cosmic gravitons we discussed explicitly two oversimplified toy
models of dimensional decoupling. Provided we keep track, in the perturbed
amplitudes, of the dependence on the internal coordinates the spectral energy
distribution will be a function of the curvature scale which the 
compactification occurs and of the typical frequency of the 
internal excitations. We found for both the models a two branches violet
spectrum.
Since the power spectra are increasing in frequency the most significant 
bounds come from small wave-length and constrain the 
curvature scale to be $H_c< 10^{-1} M_P$ (provided $\omega_{q}< \omega_c$ and
 with $\omega_c \sim 10^{11} (H_c/M_P)^{1/2} Hz $).
At the same time the violet spectra are less constrained at low  frequencies 
by the large scale observations and in particular by the $COBE$ bound. 
The estimates presented in this paper suggest that for $\omega>\omega_q$
the slopes of the spectra (but not their amplitudes) can be reliably computed 
by neglecting the internal Laplacians. On the contrary the presence 
of the internal Laplacians affects decisevely the spectral slopes 
(and amplitudes) for $\omega<\omega_q$. Our considerations 
can be also applied to more realistic models of dimensional reduction
(with the unavoidable help of numerical techniques)
in order to discuss the back-reaction problems which could eventually lead 
to the isotropisation of the original background model. It might also be of 
some interest to deepen the possible phenomenological signatures of the
scenarios of dimensional decoupling and their relevance for the formation of a 
stochastic gravity-waves background.
We want finally to stress that even though the models analyzed in this paper 
are quite simplified the perturbative techniques which we introduced are 
more general. 
The open problem which emerges also from our discussion is of course to 
understand if a viable multidimensional cosmological model, free of the 
well known problems mentioned in the introduction, exists at all. String 
theory seem to be a very good candidate for this purpose and it is very 
tempting to   speculate that the same mechanisms leading, in principle,
to a graceful exit in four dimensions\cite{18} could also operate in order 
to stabilize the internal dimensions producing, ultimately a completely
isotropic universe. It could also be possible that classical field
configurations (like Dirac monopole configurations polarized along 
the internal dimensions) can offer suitable mechanisms for the 
stabilization of the internal space \cite{9b1G} and in  
this directions the work is still in progress \cite{22c}.

\section*{Acknowledgments}

I would like to thank M. Gasperini and G. Veneziano for many discussions
and for drawing originally my attention on the subject of this investigation.
This work was supported in part the ``Human Capital and Mobility Program"  of
the European Commission, under the CEE contracts No. CHRX-CT94-0423.

\end{document}